\documentclass{egpubl}
\usepackage{vcbm2025s}

 \WsShortPaper      

\usepackage[T1]{fontenc}
\usepackage{dfadobe}  

\usepackage{cite}  
\BibtexOrBiblatex
\electronicVersion
\PrintedOrElectronic
\ifpdf \usepackage[pdftex]{graphicx} \pdfcompresslevel=9
\else \usepackage[dvips]{graphicx} \fi

\usepackage{egweblnk} 

\usepackage{enumitem}
\newcommand{\task}[1]{\mbox{$T_{#1}$}}

\title[Visual Analysis of Time-Dependent Observables]%
      {Visual Analysis of Time-Dependent Observables\\in Cell Signaling Simulations}

\author[L. Cibulski, F. Haack, A. Uhrmacher \& S. Bruckner]
{\parbox{\textwidth}{\centering L. Cibulski$^{1}$, F. Haack$^{2}$, A. Uhrmacher$^{2}$ and S. Bruckner$^{1}$ 
        }
        \\
{\parbox{\textwidth}{\centering $^1$Chair of Visual Analytics, University of Rostock\\
         $^2$Chair of Modeling and Simulation, University of Rostock
       }
}
}

\begin{document}

\teaser{
    \centering
    \includegraphics[width=.96\linewidth]{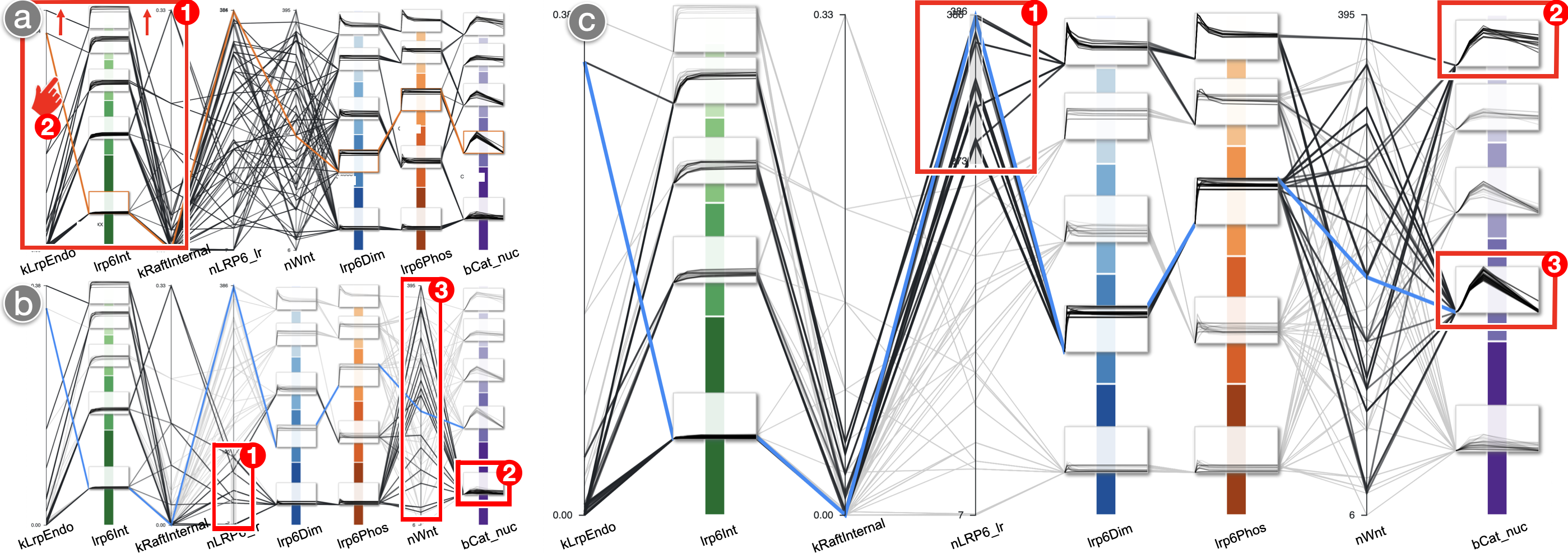}
    \caption{We embed time series as trend figures into parallel coordinates to analyze the impact of receptor trafficking on simulated signal responses over time. a) Verification: with higher rates of receptor internalization, more receptors are in fact internalized (a1), except for one outlier (a2). b) Calibration: few receptors in rafts (b1) lead to undesired low pathway activation (b2), independent of signal intensity (b3). c) Sensitivity analysis: in turn, most receptors in rafts (c1) lead to significant pathway activation (c2, c3) but only one stable behavior (c2).}
    \label{fig:teaser}
}

\maketitle
\begin{abstract}
   The ability of a cell to communicate with its environment is essential for key cellular functions like replication, metabolism, or cell fate decisions. 
   The involved molecular mechanisms are highly dynamic and difficult to capture experimentally. 
   Simulation studies offer a valuable means for exploring and predicting how cell signaling processes unfold. 
   We present a design study on the visual analysis of such studies to support 1) modelers in calibrating model parameters such that the simulated signal responses over time reflect reference behavior from cell biology research and 2) cell biologists in exploring the influence of receptor trafficking on the efficiency of signal transmission within the cell.
   We embed time series plots into parallel coordinates to enable a simultaneous analysis of model parameters and temporal outputs.
   A usage scenario illustrates how our approach assists with typical tasks such as assessing the plausibility of temporal outputs or their sensitivity across model configurations.


\end{abstract}  

\section{Introduction}
The ability of a cell to understand and communicate with its environment is at the core of all living organisms, whether single-celled or multicellular.
Cell signaling describes the mechanism by which a cell collects information from its environment, other cells, or even itself, and then responds with an action.
Many key cellular functions depend on signaling to happen correctly, e.g., replication, metabolism, or cell fate decisions, driving higher-level processes like tissue regeneration.
Anomalous signaling can be involved in human cancers or developmental disorders \cite{CN12}.

Signaling processes often involve highly dynamic molecular mechanisms that may be difficult to capture experimentally. Computational modeling, especially through simulation studies, offers detailed control, repeatability, and close observation, making it a powerful alternative for exploring and predicting how signaling processes unfold.
Researchers studying cell signaling can particularly benefit from mechanistic models, which -- unlike statistical or AI-driven models -- reflect the actual structure and behavior of biological mechanisms known to govern the phenomenon of interest. 
While advanced computational approaches to calibrating model parameters exist, expert knowledge is still crucial, e.g., to specify initial parameter values or boundary conditions, or to decide whether simulation results are plausible given the biological system at hand. 

In this paper, we present a visual analytics approach to support the exploration of simulation data in cell signaling studies.
We first outline the workflow used in simulation-based modeling of biological processes and highlight key challenges, particularly those related to the interpretation of time-dependent observables and the high dimensionality of parameter spaces.
We then propose to embed time series plots into parallel coordinates, allowing model parameters and temporal outputs to be viewed in a single representation. 
Finally, we illustrate how this approach can assist with typical analysis tasks, including plausibility checks of model behaviors, the refinement of parameter choices, and the examination of sensitivity and variation across simulation runs. 
Our goal is to support domain experts and modelers in their collaborative efforts to develop, refine, and apply mechanistic models of cell signaling processes.

\section{Related Work}

Dynamic processes are fundamental to understanding physiology across all scales~\cite{Garrison-2022-TOV}. At the organ and system level, visualizations often focus on multivariate time-dependent measurements, such as vital signs or brain activity. For example, tiled parallel coordinate plots have supported the exploration of temporal patterns in EEG recordings~\cite{tenCaat-2005-TPC}. Stoppel et al.~\cite{Stoppel-2016-GIR} integrate time curves into spatial representations to aid the interpretation of time-dependent imaging data.
At the pathway level, systems like Minardo~\cite{Ma-2015-VAS} embed time profiles into biological networks, enabling the interpretation of temporal activity in relation to known cellular mechanisms.
At the cellular scale, experimental observation becomes increasingly difficult, and simulations play an important role in hypothesis generation and model refinement. 
These simulations often involve numerous parameters and produce complex time-dependent outputs. 
Visual parameter space analysis has been proposed as a conceptual framework for exploring input-output relationships in simulation-based studies~\cite{Sedlmair-2014-VPS}, and successfully employed in a variety of domains. 
For instance, Diehl et al.~\cite{Diehl-2015-VAS} visualize temporal weather simulation data through a combination of small multiples, map-based views, and temporal brushing to explore patterns across forecast ensembles. 
Konyha et al.~\cite{Konyha-2006-IVA} explore families of function graphs using interactive brushing and linking to relate scalar input parameters to trends in temporal outputs, emphasizing analyst-driven hypothesis generation. 
Eichner et al.~\cite{Eichner-2020-MPD} analyze parameter dependencies of time-series segmentation results through several linked views: a triangular subrange correlation view, a parallel coordinates view for average correlation strength, and a tabular view for deviations from the average correlations.
In cellular simulation, tools such as those by Schulz et al.~\cite{Schulz-2011-VAS} and Luboschik et al.~\cite{Luboschik-2014-SIV} offer linked parameter and output views, but typically separate spatial, temporal, and input analyses. 

While these approaches are powerful, our work focuses on a specific need: treating distinct temporal behaviors as discrete pattern types, similar to categorical outcomes, and enabling researchers to directly relate these to multiple simulation input parameters through an integrated visualization.


\section{Data and Task Analysis}
\label{sec:problem-characterization}

Simulation models in cell signaling aim at capturing dynamic molecular mechanisms that take place around or inside a cell. Cells have various intricate ways to control the intensity of their response to external stimuli (e.g., hormones, neurotransmitters, or growth factors). 
One important way is to manage the availability and responsiveness of receptors at the cell surface through their sorting, internalization, degradation, and recycling. 
This is known as receptor trafficking. 
It is a key mechanism to dampen signal intensity by lowering the number of receptors and ligands, while also being able to promote signaling, depending on receptor state and timing. 

As a case study, we consider a computational model of membrane dynamics during the initial stages of canonical Wnt signaling \cite{HB2020}.
The cellular response follows upon the formation of a receptor complex in specialized parts of the membrane after binding of the Wnt ligand. 
Once the receptor complex is formed, an inhibitor protein is recruited from the inner cell to the membrane where it is bound and inactivated.
This initiates the intracellular signal transmission. 
Receptor sorting and internalization play a pivotal role in this process \cite{CK2021}.
Depending on the membrane structure a receptor is residing in, either raft or non-raft domain, different signaling pathways might be activated. 
If internalization is induced in non-raft domains, receptor and Wnt ligand are degraded and the signal is attenuated.
If it is induced in raft domains, a particular protein accumulates in the inner cell and the cell response is promoted. 
However, which internalization pathway is activated under which conditions is still under debate \cite{CK2021}.

In the remainder of this section, we outline the analysis tasks that arise from simulation data of such dynamic signaling processes.

\begin{figure}[tb!]
    \centering
    \includegraphics[width=\linewidth]{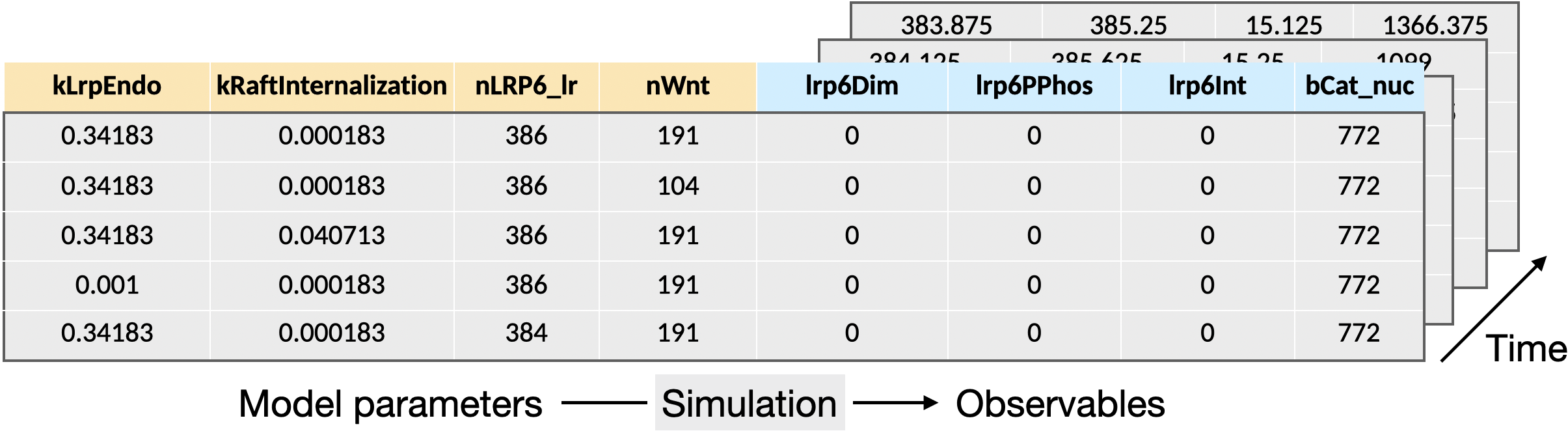}
    \caption{The simulation data includes independent model parameters (orange) and simulated time-dependent observables (blue).
    }
    \label{fig:data-structure}
\end{figure}

\noindent \textbf{Data Abstraction}\quad
The simulation model can be considered an input-output model that approximates the cell signaling behavior as a function $X\rightarrow{Y}$ by mapping some input dimensions $X=\{X_1,...,X_n\}$ to a number of output dimensions $Y=\{Y_1,...,Y_m\}$ (Figure \ref{fig:data-structure}).
We refer to the input dimensions $X$ as \emph{model parameters} and to the dependent output dimensions $Y$ as \emph{observables}.
We refer to the union $(\vec{x},\vec{y})$ of a model configuration $\vec{x}=(x_1, ..., x_n);x_i\in{X_i}$ and its observables $\vec{y}=(y_1, ..., y_m); y_i \in{Y_i}$ as a \emph{simulation run}.
Observables can have a single scalar value $y_i$ for each $\vec{x}$. 
However, they can also be time-dependent, where $y_i$ is in fact a time series $f(\vec{x},t)$.

For time-dependent observables, \emph{required} behavior is often known from domain knowledge, previously validated models, or laboratory experiments. 
For example, the percentage of receptors that carry a ligand is known to change over time in a defined way or to overcome a certain threshold within a given time range. 

\noindent \textbf{Analysis Tasks}\quad
The challenge of such simulation studies lies in the absence of a direct inverse relation $Y \rightarrow X$ that would allow us to compute the model configuration $\vec{x}$ that results in the prescribed temporal behaviors $y_i$.
A parameter space analysis involving temporal observables might thus be guided by three main questions:

\begin{enumerate}[leftmargin=11pt]
    \item How to choose the model parameters for the temporal observables to reflect the behavior that was measured experimentally in the laboratory? For example, one might want to identify the exact rates at which the receptor trafficking mechanisms occur. 
\end{enumerate}

Once suitable model parameters have been identified, follow-up analyses can advance the understanding of cell signaling:

\begin{enumerate}[resume,leftmargin=11pt]
    \item How much can a condition be varied before a temporal observable changes into a completely different behavior? For example, one might want to understand how much the external signal intensity can vary before the transmission of the signal breaks.
    \item Which model parameter shows little variation for a regular condition and significant variation for anomalous conditions? For example, one might want to identify a mechanism to target with medication to positively influence dysfunctional cell signaling without disturbing healthy processes.
\end{enumerate}

From the data abstraction and guiding questions, we have derived the following high-level tasks to inform the visual design:

\makeatletter
\newcommand{\taskitem}[1][]{
  \protected@edef\@currentlabel{#1}%
\item[#1]
}
\makeatother
\begin{enumerate}[ref=\theenumi, wide,labelindent=0pt]
\taskitem[\task{1}]\label{task:validation} \textbf{Verification} -- \emph{Does the simulation produce plausible time series behaviors?} Implausible behaviors hint at (structural) problems in the model. For example, from previous research in cell biology, we know that the coupling of receptors, receptor activation, and pathway activation take place one after the other. If the time-dependent observable for receptor activation (\emph{lrp6Phos}), however, shows a response (e.g., a peak) before the observable for receptor coupling (\emph{lrp6Dim}) does, we can deduce that the simulation model does not reflect the correct execution order of these processes. 
\taskitem[\task{2}]\label{task:regions-of-interest} \textbf{Calibration} -- \emph{Which configurations are associated with required or expected time series behaviors?} 
The initial model parameter sampling might be sub-optimal due to many degrees of freedom and interval recommendations informed by previous findings in cell biology being available only for some parameters.
Even sophisticated parameter estimation methods can hardly cover the huge search space.
Time series behaviors as prescribed by the requirements can help identify promising model parameter ranges. 
They hint at regions of interest in the model parameter space that already capture the targeted system behavior quite well and could be further exploited with a refined sampling.
If only a fraction of simulation runs show required time series behavior at all, the model parameter sampling might need more drastic adjustments. 
\taskitem[\task{3}]\label{task:outliers} \textbf{Outlier Analysis} -- \emph{Which configurations result in outlier behaviors?} An outlier can be identified based on the data only or from expectations based on experience and domain knowledge. In contrast to implausible behavior (\ref{task:validation}), outliers might represent properties of the cell signaling process that are valid in principle. Detailed analysis is needed to distinguish between an outlier representing 1) an emergent phenomenon that has not been observed before and thus might inspire new findings (as opposed to locating what is already known in \ref{task:regions-of-interest}) and 2) a problem in the model (compare \ref{task:validation}). 
\taskitem[\task{4}]\label{task:sensitivity-analysis} \textbf{Sensitivity Analysis} -- \emph{How does the behavior of temporal observables change when model parameters change?} 
The sensitivity of time series behavior towards model parameter changes conveys how these parameters influence the behaviors of observables. 
Current assessments of sensitivity as scalar correlation values might not capture all variation patterns, in particular of time series. 
In contrast to \ref{task:regions-of-interest} and \ref{task:outliers}, which aim to identify parameter ranges associated with certain behaviors, this task aims to detail the exact way, in which changes of model parameter values induce changes in temporal behaviors. 
Visual sensitivity analysis helps steer parameter variation based on domain-specific notions of (dis-)similar temporal behaviors to arrive at the required behavior.
It also conveys a notion of the robustness of an observed behavior, e.g., analysts are often interested in tipping points of the temporal developments.
\end{enumerate}


\section{Temporal Parallel Coordinates Axes}

Gaining insight into the correspondence between model configurations and time-dependent observables requires a lossless two-dimensional visual representation for multi-dimensional simulation runs. 
Parallel coordinates are a widely used and well-studied technique for the visualization of multivariate data, in particular when it comes to simulation data \cite{HW13}.
They offer flexible axis layouts and support a variety of tasks in different application domains including life sciences and cell biology in particular \cite{HW13}.
We build upon an open-source implementation\footnote{PAVED has been developed at Fraunhofer IGD and is available at \url{https://github.com/fraunhofer-igd-iva/pavedjs}.} of parallel coordinates called PAVED \cite{CMMK20}, which has been successfully applied to a conceptually similar use case on multi-attribute decision-making \cite{CM24}.
PAVED offers a compact overview of multivariate simulation runs and simple interactions for selecting runs with desired behavior and eliminating runs with undesired behavior.
However, it has not been designed to handle time-dependent outputs.

Parallel coordinates can be extended in different ways to include the time dimension. 
One option is to depict time on an additional axis \cite{WLG97, TAS04}, but this often leads to overplotting and fails to preserve the continuity of individual time series. 
Aggregating time series to scalar values \cite{DKG12} may obscure important temporal dynamics. 
Representing time steps as separate axes \cite{Edsall03} does not scale well with many observables or time points. 

We instead aim for a simultaneous depiction of model parameters and time-dependent observables in a unified view, treating time series as complex data objects.
To do this, we build upon the concept of minimalistic trend figures \cite{ZLTS03}, but depict the changes of observable values over time rather than across data items. For each simulation run and observable, one trend figure is visually aligned with the corresponding polyline in the parameter space. In contrast to approaches using vanishing point perspectives \cite{GRPF16}, we avoid perceptual distortion by following the concept of juxtaposed nested plots \cite{WLSL16}.

To reduce visual clutter and highlight key patterns, we cluster similar time series for each observable. 
This reflects the common situation that only a few distinct behaviors are of interest to domain experts. 
We use k-means clustering with dynamic time warping to group similar temporal profiles and determine the number of clusters empirically based on expert feedback. 
Since clustering is not always fully reliable and might not capture domain-specific notions of similarity, we enable interaction with the clustered output to support visual verification and refinement by users.

Similar as in parallel sets \cite{KBH06}, we replace the point intersections at the axes with boxes that represent the clusters.
These boxes are scaled according to the size of the clusters on that axis (i.e., the number of cluster members) relative to all data samples (i.e., model configurations).
Their initial ordering also follows the cluster size, with larger clusters being at the bottom of an axis and smaller clusters being at the top.
Color is used to differentiate the clusters, with hue indicating the time-dependent attribute the cluster belongs to and saturation indicating the different clusters within that attribute, again reflecting their sizes (clusters with more members are rendered with more saturation). 
Each colored box is overlayed with a multi-line chart that superimposes the members of the respective time series cluster.
The x-axis reflects time and the y-axis reflects the attribute values.
The time range is identical across all observables.
The y-axis is scaled according to the minimum and maximum values of the respective attribute across all time series.
Axis ticks and time point marks are omitted to save screen space and focus on temporal trends.

We extend the interaction of the non-temporal PAVED implementation \cite{CMMK20} to our newly introduced temporal axes, such that time series are highlighted or grayed out in accordance with their respective polyline.
Consequently, we support several common interaction schemes that allow analysts to filter model configurations according to constraints and preferences, browse the resulting selection, highlight and bookmark model configurations or temporal behaviors of interest, inspect details on demand, reorder the axes and clusters according to their relevance to the task, and refine the clustering according to perceived similarity of behaviors.

\section{Usage Scenario}

We illustrate the effectiveness of our visual analytics approach by applying it to a simulation model of receptor internalization during the initial stages of canonical Wnt signaling.
The simulation model is defined using the modeling language \emph{ML-Rules} \cite{HWMU2017}, where rules describe the signaling cascade together with known cellular dynamics, e.g., binding kinetics or the activation of receptor complexes, that are obtained from experimental measurements or previously validated models.
A parameter scan of the simulation model is then executed with a stochastic simulation engine \cite{KHWU2024}.

By analyzing the parameter scan (Figure \ref{fig:data-structure}), our collaborators aim to contribute to the ongoing debate about the exact mechanism of signaling pathway activation (Section \ref{sec:problem-characterization}).
They systematically investigate the hypothesized influence of the membrane structures on the temporal regulation of receptor coupling, internalization, activation, and ultimately pathway activation under different conditions.
These conditions, 141 in total, are reflected by varying the model parameter values: the intensity of the external Wnt stimuli (\emph{nWnt}), the receptor distribution between the two membrane structures (proportion \emph{nLRP6\_lr} of receptors in raft domains), and the rate of receptor internalization in each of these structures (\emph{kRaftInternal} for raft domains and \emph{kLrpEndo} for non-raft domains).
Based on the analysis aim, the following observables serve as the model output: the number of coupled receptors (\emph{lrp6Dim}), the number of internalized receptors (\emph{lrp6Int}), the number of active receptors (\emph{lrp6Phos}), and the accumulation of the beta-catenin protein (\emph{bCat\_nuc}) as a measurement of pathway activation. 
As external stimuli can trigger immediate but also longer-term responses in the cells, seven time steps are observed: 0, 10, 20, 30, 60, 120, and 360 minutes after stimulation.

If users need to verify that the internalization response in the simulated signaling behavior actually depends on the responsible model input (\ref{task:validation}), they can use drag and drop to move the response axis between the axes representing the model parameters that control internalization. 
In this case, their expectation that increasing values of the internalization parameters (\emph{kLrpEndo} and \emph{kRaftInternal}) result in increasing numbers of receptors being internalized over time (\emph{lrp6Int}) is confirmed (Figure \ref{fig:teaser}a1). 
However, one configuration does not follow this pattern: a low \emph{lrp6Int} response is observed despite a high value of the \emph{kLrpEndo} parameter (Figure \ref{fig:teaser}a2). 
Closer inspection of the polyline of this semantic outlier (\ref{task:outliers}) reveals that it comprises the highest value of the \emph{nLRP6\_lr} parameter, i.e., most receptors are located in raft domains. 
Receptors in raft domains, however, are not available for the internalization pathway represented by the \emph{kLrpEndo} parameter.
Thus, the high \emph{kLrpEndo} value has a diminishing effect on the model outcome, as it applies only to a small minority of receptors. 

Given a verified model, brushing may be used to explore the influence of different parameters.
For instance, users interested in the impact of receptor localization (\emph{nLRP6\_lr}) on the signal activation (\emph{bCat\_nuc}), can start by brushing configurations where most receptors are in non-raft domains (Figure \ref{fig:teaser}b1). 
The selection reveals consistently low signal activation (\emph{bCat\_nuc}, Figure \ref{fig:teaser}b2), independent of the stimuli intensity (\emph{nWnt}, Figure \ref{fig:teaser}b3). 
This observation reproduces previous findings reported in literature \cite{CK2021}.

If users aim to understand how the signal activation changes when receptors are mostly in raft domains (\ref{task:sensitivity-analysis}), they can move the brush from low to high values (Figure \ref{fig:teaser}c1). 
While this reveals two clusters with high signal activation, our collaborators unexpectedly find that only one cluster reproduces a stable signal activation (Figure \ref{fig:teaser}c2). 
The second cluster reveals undesired transient signal activation, in which the \emph{bCat\_nuc} trajectory declines to its ground level after an initial peak (Figure \ref{fig:teaser}c3). 
They suspect that this relationship is a previously non-described phenomenon in cell biology that warrants further research.
Hovering or brushing the respective clusters (\ref{task:regions-of-interest}) can help further distinguish these two behaviors by comparing the corresponding regions of the model parameter space, potentially identifying influencing factors to be investigated in the future. 

\section{Discussion and Future Work}

The usage scenario and feedback from our domain experts suggest a significant value of our visual analysis approach for simulation-based understanding of cell signaling processes.
With simulation models in computational biology being quite large, hundreds to thousands of simulation runs are possible.
While this does not directly influence the number of time series clusters, techniques from time-oriented visualization research might be needed to cope with increasing numbers of cluster members in the multi-line charts.
When simulating cell signaling processes, we can assume the time series to show an initial response (e.g., a peak or increase/decrease) and then converge into some kind of stable behavior. A generalization of the multi-line charts to more fluctuating or oscillating time series beyond cell signaling simulations remains an open challenge.

Future work could dive deeper into task-dependent representations of temporal value changes at the parallel coordinates axes and introduce dedicated interactions to seamlessly switch between different granularities and contexts, e.g., moving from identifying representative behaviors in clusters over analyzing one selected clusters to a detailed inspection of a selected time series -- and back.
Our domain experts also commented that their simulation study workflow would significantly benefit from being able to directly execute additional runs from our visualization tool based on interesting model configurations or interesting temporal developments observed in the model output.
This request for simulation steering also motivates a more progressive analysis approach.

\section{Conclusion}

We presented a design study on simultaneously exploring model parameters and temporal outputs to support the simulation-based modeling and prediction of how cell signaling processes unfold.
Our interactive visualization embeds time series plots into parallel coordinates to address key challenges related to the identification of (im)plausible temporal outputs, the calibration of model parameters, and the examination of how temporal outputs vary in response to parameter refinements.
Our usage scenario illustrates how our approach can assist in understanding  signaling pathway activation and our collaborators in computational biology expressed strong interest in using our tool for their research and developing it further.

\bibliographystyle{eg-alpha}

\bibliography{references}

\newcommand{\etalchar}[1]{$^{#1}$}
\begin{thebibliography}{\uppercase{CMMK20}}

\bibitem[CK21]{CK2021}
\textsc{Colozza G., Koo B.-K.}:
\newblock Wnt/{$\beta$}-catenin signaling: Structure, assembly and endocytosis
  of the signalosome.
\newblock \emph{Dev. Growth Differ. 63}, 3 (2021), 199--218.

\bibitem[CM24]{CM24}
\textsc{Cibulski L., May T.}:
\newblock Revisiting {PAVED}: Studying tool adoption after four years.
\newblock In \emph{Proc. EuroVis -- Short Papers} (2024).

\bibitem[CMMK20]{CMMK20}
\textsc{Cibulski L., Mitterhofer H., May T., Kohlhammer J.}:
\newblock {PAVED}: Pareto front visualization for engineering design.
\newblock \emph{Comput. Graph. Forum 39}, 3 (2020), 405--416.

\bibitem[CN12]{CN12}
\textsc{Clevers H., Nusse R.}:
\newblock Wnt/$\beta$-catenin signaling and disease.
\newblock \emph{Cell 149}, 6 (2012), 1192--1205.

\bibitem[DKG12]{DKG12}
\textsc{Dasgupta A., Kosara R., Gosink L.}:
\newblock Meta parallel coordinates for visualizing features in large,
  high-dimensional, time-varying data.
\newblock In \emph{Proc. IEEE LDAV} (2012), pp.~85--89.

\bibitem[DPD{\etalchar{*}}15]{Diehl-2015-VAS}
\textsc{Diehl A., Pelorosso L., Delrieux C., Saulo C., Ruiz J., Gr{\"o}ller
  M.~E., Bruckner S.}:
\newblock Visual analysis of spatio-temporal data: Applications in weather
  forecasting.
\newblock \emph{Comput. Graph. Forum 34}, 3 (2015), 381--390.

\bibitem[Eds03]{Edsall03}
\textsc{Edsall R.~M.}:
\newblock The parallel coordinate plot in action: Design and use for geographic
  visualization.
\newblock \emph{Comput. Stat. Data Anal. 43}, 4 (2003), 605--619.

\bibitem[EST20]{Eichner-2020-MPD}
\textsc{Eichner C., Schumann H., Tominski C.}:
\newblock Making parameter dependencies of time-series segmentation visually
  understandable.
\newblock \emph{Comput. Graph. Forum 39}, 1 (2020), 607--622.

\bibitem[GKV{\etalchar{*}}22]{Garrison-2022-TOV}
\textsc{Garrison L.~A., Kolesar I., Viola I., Hauser H., Bruckner S.}:
\newblock Trends \& opportunities in visualization for physiology: A multiscale
  overview.
\newblock \emph{Comput. Graph. Forum 41}, 3 (2022), 609--643.

\bibitem[GRPF16]{GRPF16}
\textsc{Gruendl H., Riehmann P., Pausch Y., Froehlich B.}:
\newblock Time-series plots integrated in parallel-coordinates displays.
\newblock \emph{Comput. Graph. Forum 35}, 3 (2016), 321--330.

\bibitem[HBU20]{HB2020}
\textsc{Haack F., Budde K., Uhrmacher A.~M.}:
\newblock Exploring the mechanistic and temporal regulation of lrp6 endocytosis
  in canonical wnt signaling.
\newblock \emph{J. Cell Sci. 133}, 15 (2020), jcs243675.

\bibitem[HW13]{HW13}
\textsc{Heinrich J., Weiskopf D.}:
\newblock State of the art of parallel coordinates.
\newblock \emph{Proc. Eurographics -- STARs} (2013), 95--116.

\bibitem[HWMU]{HWMU2017}
\textsc{Helms T., Warnke T., Maus C., Uhrmacher A.~M.}:
\newblock Semantics and efficient simulation algorithms of an expressive
  multilevel modeling language.
\newblock \emph{ACM Trans. Model. Comput. Simul. 27}, 2, 8:1--8:25.

\bibitem[KBH06]{KBH06}
\textsc{Kosara R., Bendix F., Hauser H.}:
\newblock Parallel sets: Interactive exploration and visual analysis of
  categorical data.
\newblock \emph{IEEE Trans. Vis. Comput. Graph. 12}, 4 (2006), 558--568.

\bibitem[KHWU]{KHWU2024}
\textsc{K\"oster T., Henning P., Warnke T., Uhrmacher A.}:
\newblock Expressive rule-based modeling and fast simulation for dynamic
  compartments.
\newblock \emph{PLOS ONE 19}, 10, e0312813.

\bibitem[KMG{\etalchar{*}}06]{Konyha-2006-IVA}
\textsc{Konyha Z., Matkovic K., Gracanin D., Jelovic M., Hauser H.}:
\newblock Interactive visual analysis of families of function graphs.
\newblock \emph{IEEE Trans. Vis. Comput. Graph. 12}, 6 (2006), 1373--1385.

\bibitem[LRHS14]{Luboschik-2014-SIV}
\textsc{Luboschik M., Rybacki S., Haack F., Schulz H.-J.}:
\newblock Supporting the integrated visual analysis of input parameters and
  simulation trajectories.
\newblock \emph{Comput. Graph. 39} (2014), 37--47.

\bibitem[MSK{\etalchar{*}}15]{Ma-2015-VAS}
\textsc{Ma D. K.~G., Stolte C., Kaur S., Bain M., O'Donoghue S.~I.}:
\newblock Visual analytics of signalling pathways using time profiles.
\newblock \emph{Adv. Exp. Med. Biol. 823} (2015), 3--22.

\bibitem[SHB{\etalchar{*}}14]{Sedlmair-2014-VPS}
\textsc{Sedlmair M., Heinzl C., Bruckner S., Piringer H., M{\"o}ller T.}:
\newblock Visual parameter space analysis: A conceptual framework.
\newblock \emph{IEEE Trans. Vis. Comput. Graph. 20}, 12 (2014), 2161--2170.

\bibitem[SHHB16]{Stoppel-2016-GIR}
\textsc{Stoppel S., Hodneland E., Hauser H., Bruckner S.}:
\newblock Graxels: Information rich primitives for the visualization of
  time-dependent spatial data.
\newblock In \emph{Proc. EG VCBM} (2016), pp.~183--192.

\bibitem[SUS11]{Schulz-2011-VAS}
\textsc{Schulz H.-J., Uhrmacher A.~M., Schumann H.}:
\newblock Visual analytics for stochastic simulation in cell biology.
\newblock In \emph{Proc. I-Know} (2011), pp.~48:1--48:8.

\bibitem[TAS04]{TAS04}
\textsc{Tominski C., Abello J., Schumann H.}:
\newblock Axes-based visualizations with radial layouts.
\newblock In \emph{Proc. ACM SAC} (2004), pp.~1242--1247.

\bibitem[tCMR05]{tenCaat-2005-TPC}
\textsc{ten Caat M., Maurits N.~M., Roerdink J. B. T.~M.}:
\newblock Tiled parallel coordinates for the visualization of time-varying
  multichannel {EEG} data.
\newblock In \emph{Proc. EuroVis} (2005), pp.~61--68.

\bibitem[WLG97]{WLG97}
\textsc{Wegenkittl R., L\"{o}ffelmann H., Gr\"{o}ller E.}:
\newblock Visualizing the behaviour of higher dimensional dynamical systems.
\newblock In \emph{Proc. IEEE Visualization} (1997), pp.~119--125.

\bibitem[WLSL16]{WLSL16}
\textsc{Wang J., Liu X., Shen H.-W., Lin G.}:
\newblock Multi-resolution climate ensemble parameter analysis with nested
  parallel coordinates plots.
\newblock \emph{IEEE Trans. Vis. Comput. Graph. 23}, 1 (2016), 81--90.

\bibitem[ZLTS03]{ZLTS03}
\textsc{Zhao K., Liu B., Tirpak T.~M., Schaller A.}:
\newblock Detecting patterns of change using enhanced parallel coordinates
  visualization.
\newblock In \emph{Proc. IEEE ICDM} (2003), pp.~747--750.

\end{thebibliography}

\end{document}